\documentclass[12pt]{article}

\newcommand{\ssf}[1]{\mbox{\small \rm #1}}

\newcommand{\dtm}[1]{\ssf{Det}\left(#1\right)}
\newcommand{\diag}[1]{\ssf{diag}\left(#1\right)}
\newcommand{\rank}[1]{\ssf{rank}\left(#1\right)}

\newcommand{\cnt}[1]{
\begin{center}#1
\end{center}}

\newcommand{\spc}{\hspace*{5mm}}

\newcommand{\et}{
\hspace{5mm}\mbox{and}
\hspace{5mm}}

\newcommand{\ou}{
\hspace{5mm}\mbox{where}
\hspace{5mm}}

\newcommand{\dm}[2]{\mbox{\footnotesize $ \,[{#1} \times {#2}] $}}

\newcommand{\rf}[1] {{\bf(\protect\ref{#1})}}

\newcommand{
\beq}[1] {
\begin{equation} \label{#1} }

\newcommand{ \eeq
}{
\end{equation} }

\newcommand{\btb}[3] { \protect
\begin{table}[hbt]
\begin{center} \protect\caption{#1} \label{#2} \vspace{1ex}
\small
\begin{tabular} {#3} }

\newcommand{\etb}{
\end{tabular}
\end{center}
\end{table} }

\newcommand{\upb}[2]{
\begin{array}c{#1} \\
{#2}
\end{array}}

\newcommand{\upa}[1]{\upb{#1}{}}

\newcommand{\upc}[3]{\upb{#1}{\dm{#2}{#3}}}

\usepackage{graphicx}
\usepackage{amsmath,amsthm,amscd,amssymb,eufrak,latexsym,upref}

\newcommand{\mydate}{Feb. 2, 2013 @10:37}

\newcommand{\itt}[1]{

\item{{\bf \em #1} \\
}}

\newcommand{
\skp}[1] {}

\newcommand{
\tmpry}[1] {}

\newcommand{\der}[1]{\frac{\partial}{\partial{#1}}}

\newcommand{\fixp}{\ssf{FixedPoint}}

\newcommand{\sdl}{\ssf{SaddlePoint}}

\newcommand{\svd}{\ssf{SVD}}

\newcommand{\rms}{\ssf{Rmse}}

\newcommand{\cut}[2]{\ssf{Cut}_{(#1,#2)}}

\newcommand{\miin}[1]{\ssf{Min}_{#1}}

\newcommand{\apx}{\ssf{WLRA}}

\newcommand{\apxx}[2]{\apx_{#1 \tau=#2}}

\newcommand{\mnz}[1]{\parallel{#1}\parallel}

\begin{document}
\[ \]

\cnt{{\bf{\Large On Weighted Low-Rank Approximation} \\
\small William Rey\footnote{Dr. Rey (widdy.rey(at)gmail.com) is
retired from the Philips Research Laboratories, Eindhoven, The
Netherlands.}}}

\begin{abstract}{\bf \sf Our main interest is the low-rank approximation
of a matrix in $ {\mathbb R}^ {m \times n} $ under a weighted
Frobenius norm. This norm associates a weight to each of the $
(m \times n) $ matrix entries. We conjecture that the number of
approximations is at most $\min (m, n) $. \\
We also investigate how the approximations depend on the weight-values.

\mbox{} \\
Keywords:} Weight, Low-rank, Factorization, Missing.
\end{abstract}

This is a short investigation concerning a best approximation
of an arbitrary real matrix $X$ by a ``weighted" low-rank approximation
$ \apx$ according to
\[ \upc X m n \upa {\approx} \upc A m p \upc {B',} p n \upa = \upc
{\apx} m n \]
with a condition on the ranks
\[ \rank A = \rank B = p \et p < \rank X. \]
All matrices are real with $X \in {\mathbb R}^ {m \times n} $,
$A \in {\mathbb R}^ {m \times p} $ and $B \in {\mathbb R}^ {n
\times p} $.

The attention is more focused on the approximations $ \apx$ than
on the factors $A$ and $B$. The approximations under consideration
minimize the weighted Frobenius norm
\beq {eq1} \parallel X - A \; B'\parallel_{w^2}^2 = \sum_{i
= 1}^m \sum_{j = 1}^n w_{i, j}^2 \; \left[X_{i, j} - (A
\; B')_ {i, j} \right]^2 = \miin {A, B}, \eeq
as also defined by (1.3) of Higham (2002).

In some sense, the minimization we face goes much further than
the indefinite least squares problem of Bojanczyk, Higham and
Patel (2003). The fact that both factors $A$ and $B$ are of concern
and the allocation of a separate weight to each of the $X$-entries
seriously complicate the matter. Nevertheless, this structure
naturally arises in some contexts such as the smoothing of a data
matrix of measurements where each of the measured entity can have
its own limited precision. The scientists working in this domain
know that minimization \rf{eq1} can have several solutions, a
badly understood feature that can loosely be attributed to the
lack of convexity of this norm in terms of all the entries of
$A$ and $B$, although the norm is convex in the entries of $ (A
\, | \, \mbox {given} \; B) $ and those of $ (B \, | \, \mbox
{given} \; A) $.

What is the possible number of best approximations $ \apx = A
\; B'$, under the weighted Frobenius norm  \rf{eq1}? \\
This minimization is a NP-hard problem, as shown by Gillis and
Glineur (2010). What exactly takes place is so little known that
it may explain why many authors of application papers ignore the
fact. Since Bradu and Gabriel (1978) introduced the biplot, many
variations on approximating by a low-rank matrix have been investigated
(Greenacre, 2012).

The paper leads to a conjecture on the number of solutions; they
can be quite many. Having started with generalities, Section 2
addresses some aspects of algorithmic convergence. Then, we turn
our attention to a dual problem where we consider that the weights
can vary, while the matrix entries are seen as arbitrary constants.
At Section 4, we even imagine that the squared weights can become
negative and wonder what can be expected from such an algebraic
trick. This is a domain where the experience is fairly limited,
although the trick gives insight on our problem. Sections 5 and
6 report numerical experiments that lead to the conjecture. Eventually,
some algorithmic notes are the object of Section 8.

\begin{enumerate}

\itt{Generalities.} There exist several algorithms for the evaluation
of $ \apx$ and the one with the lowest computational complexity
is based on two weighted linear regression steps. Starting with
an arbitrary estimate of $A$, $A^{(k)} $, it is improved by first
evaluating the corresponding optimal $B^{(k)} $,
\beq {eq2} B^{(k)} : \; \mnz {X - A^{(k)} \; B'}_ {w^2}^2
= \miin B, \eeq
and then estimating an improved $A^{(k + 1)} $,
\beq {eq3} A^{(k + 1)} : \; \mnz {X - A \; {B^{(k)}} '}_ {w^2}
^2 = \miin A, \eeq
until the limit
\[ \apx^{(\infty)} \approx A^{(k)} \; {B^{(k)}} ' \]
is sufficiently well attained. This algorithm is said to be {\em
alternating} and has been thoroughly investigated in statistics
since Gabriel and Zamir (1979); their method does not necessarily
converge to an absolute minimum point (see the discussion in Section
6, page 491).

Seeing that
\beq {eq4} A \; B' = (A \; M) \; \times \; (M^{- 1} \; B') \ou
M \; \mbox {is full rank,} \eeq
the factors $A$ and $B$ are defined up to the matrix scaling factor
$M$. Hence, clearly, we estimate too many parameters when we work
on all entries of $A$ and $B$; the feature dramatically complicates
the algorithmic test of convergence. To obviate the difficulty,
several other algorithms have been investigated ranging from brute
force minimization on all entries of $A$ (or $B$) to limiting
the space search to a Grassmann product manifold, Manton et al.
(2003). Various comparisons of the possible strategies have been
reported and Srebro and Jaakkola (2003) deserves a special mention.
Simonsson and Eld\'en (2010), Yan (2010), Markovsky (2010) as well
as Markovsky and Van Huffel (2007) cover more specific aspects.
Note that Van Huffel's school investigates a norm that is slightly
generalized compared to \rf{eq1}; their weight structure permits
to introduce correlations between the $X$-entries. The recent
work of Usevich and Markovsky (2012) about structured matrices
is worthy of attention. Maronna and Yohai (2008) pay particular
attention to algorithm initialisation. Okatani, Yoshida and Deguchi
(2011) compare several methods where a damping factor is helping
to find the global minimum solution; their study is restricted
to the field of computer vision.

\itt{On convergence and conditioning.} The two steps \rf{eq2}
and \rf{eq3} lead to convergence seeing that each of them reduces
the norm $ \mnz._{w^2}^2$. This is a guarantee of convergence,
although not of unicity of the approximation.

The iterations nicely converge inasmuch as the implied weighted
regressions are sufficiently well conditioned. Focusing on \rf{eq2},
the $j$-th column of $B'$ is evaluated by solving the equation
\[ \sum_{i = 1}^m w_{i, j}^2 \left[X_{i, j} - \left(\sum_
{k = 1}^p a_{ik} \; b_{jk} \right) \right]^2 = \miin {b_{jk}},
\spc \mbox {for} \; j = 1, 2, ..., n \]
or
\[ (X - A \; b_j) '\; W_j \; (X - A \; b_j) = \miin {b_j}, \spc
W_j = \diag {w_{1j}^2, ..., w_{mj}^2} . \]

Hence, the next $m + n$ conditions arise
\beq {eq5} \dtm {B '\; W^i \; B} \neq 0 \et \dtm {A '\; W_j \;
A} \neq 0, \eeq
where
\[ W^i = \diag {w_{i1}^2, ..., w_{in}^2} . \]
Their dependence on $A$ and $B$ is weak, seeing \rf{eq4}.

\itt{Generalisation to weights seen as variables.} The minimization
problem \rf{eq1} has for solution $ \apx = (A \; B') $ that is
such that the derivatives in its vicinity cancel. Hence, it can
also be viewed as a `saddle point' as well as a `fixed point'
of the mapping $A \rightarrow {A^{(\infty)}} $ in
\[ (A \; B') : \left\{
\begin {array} l {A^{(\infty)}} = A = \sdl = \fixp = \miin {A,
B} \\
\hspace {8mm} \mbox {where} \\
B = B (A) : \; \sum_{i = 1}^m \sum_{j = 1}^n w_{i, j}^2
\left[X_{i, j} - (A \; B')_ {i, j} \right]^2 = \miin B, \\
{A^{(\infty)}} = {A^{(\infty)}} (B) : \; \sum_{i = 1}^m \sum_
{j = 1}^n w_{i, j}^2 \left[X_{i, j} - ({A^{(\infty)}} \;
B')_ {i, j} \right]^2 = \miin {A^{(\infty)}} .
\end {array} \right.
\skp {. \} close} \]

This equivalence between $ \miin {A, B}, \fixp $ and $ \sdl$ results
from the convexity of $ \parallel X - A \; B'\parallel_{w^2}
^2 $ in the vicinity of the solution. Again, observe that the
convexity of \rf{eq1} is in term of product $A \; B'$ but not
jointly in both $A$ and $B$, which practically means that the
algorithms can only, if at all, guarantee convergence to a local
minimum. We will shortly loose this convexity but, for the time
being, let us be a little more precise on the sort of derivatives
we invoke when we speak of a saddle point.

We address a function $F$ of a matrix $X$; the function $F (X)
$ is in a metric space and varies continuously with variations
of $X$. We expect $F (X) $ to be G\^ateaux differentiable and, in
a broadly generalized sense, we defined a `saddle point'$X_s$
as being a point where all the directional derivatives vanish,
namely
\[ \left. {\der {X_s} F (X)} \right|_ {X = X_s} = \lim_{h \rightarrow
0} \frac {F (X_s + h \; \Delta X) - F (X_s)} h = 0, \spc \parallel
\Delta X \parallel > 0 \]
whatever the bounded perturbation $ \Delta X$ is. The present
extension of the `saddle point' concept places us at some distance
of the remarkable work of Benzi, Golub and Liesen (2005).

At the risk of loosing the convexity, we substitute pseudo-weights
$z_{i, j} $ to the non-negative $w_{i, j}^2$. Then, our problem
takes the form
\[ (A \; B') : \\
\left\{
\begin {array} l {A^{(\infty)}} = A = \fixp \\
\hspace {8mm} \mbox {where} \\
B = B (A) : \; \sum_{i = 1}^m \sum_{j = 1}^n z_{i, j} \left
[X_{i, j} - (A \; B')_ {i, j} \right]^2 = \sdl_B, \\
{A^{(\infty)}} = {A^{(\infty)}} (B) : \; \sum_{i = 1}^m \sum_
{j = 1}^n z_{i, j} \left[X_{i, j} - ({A^{(\infty)}} \; B')
_{i, j} \right]^2 = \sdl_{A^{(\infty)}}
\end {array} \right.
\skp {. \} close} \]
or rather
\beq {eq6} (A \; B') : \left\{
\begin {array} l \apx = \left({A^{(\infty)}} \; {B^{(\infty)}}
'\right) = A \; B' = \fixp \\
\hspace {8mm} \mbox {where} \\
{B^{(\infty)}} : \; \der {B^{(\infty)}} \; \sum_{i = 1}^m
\sum_{j = 1}^n z_{i, j} \left[X_{i, j} - \left(A \; {B^
{(\infty)}} '\right)_ {i, j} \right]^2 = 0, \\
{A^{(\infty)}} : \; \der {A^{(\infty)}} \; \sum_{i = 1}^m
\sum_{j = 1}^n z_{i, j} \left[X_{i, j} - \left({A^{(\infty)}}
\; {B^{(\infty)}} '\right)_ {i, j} \right]^2 = 0.
\end {array} \right.
\skp {. \} close} \eeq
The difference between the last two formulations is computational.

\itt{Path following and anticipations.} What is the possible number
of best approximations $ \apx = A \; B'$, given a set of weights?

With this question as an investigation direction, we first converted
the problem to the saddle point set-up \rf{eq6}. This induces
us to consider the solutions as being functions of the pseudo-weights
$z_{i, j} $ and we wonder how these approximations vary when
the pseudo-weights vary.

Clearly, independently varying all the pseudo-weights greatly
increases the number of concerned variables and the complexity
of the problem. In order to keep it simple and to be able to `see'
how the solutions behave, we limited ourselves to following a
``path". This is now supported with the help of a numerical example
to clarify the details.

Starting with solutions on pseudo-weights, the path is in the
space of pseudo-weights $z_{i, j} $ and is parametrized by a
parameter $ \tau$.

\begin{itemize}

\item For $ \tau = 0$, the given set of pseudo-weights $ \left\{...,
z_{i, j}, ... \right\} $ is of concern,
\[ \tau = 0 \spc \Longrightarrow \spc {\cal Z}_0 = \left\{...,
z_{i, j}, ... \right\} = \left\{..., w_{i, j}^2, ... \right
\} . \]

\item For $ \tau = 1$, the unweighted case (with a unique minimum)
is met with,
\[ \tau = 1 \spc \Longrightarrow \spc {\cal Z}_1 = \left\{\bar
z, ..., \bar z \right\} \ou \bar z = \frac {\sum_{i, j} w_{i,
j}^4} {\sum_{i, j} w_{i, j}^2} . \]

\item For $ | \tau - 1 | \gg 0$, the pseudo-weight set $ {\cal
Z}_ {\tau} $ consists of positive and negative entries. Several
saddle point solutions to \rf{eq6} can be expected.

\end{itemize}

Eventually, we follow the path from $ \tau \ll 0$ to $ \tau \gg
1$, passing for $ \tau = 0$ by our set of interest, $ {\cal Z}
_0$. Such a path can be described by
\beq {eq7} {\cal Z}_ {\tau} = {\cal Z}_0 + \tau \; ({\cal Z}
_1 - {\cal Z}_0) . \eeq
Note that more sophisticated forms could be of interest.

Bearing our attention on the solutions of the saddle point problem
\rf{eq6} while following this path,

\begin{itemize}

\item we know that
\[ \tau = 1 \spc \Longrightarrow \spc \mbox {a unique solution}
. \]

\item By continuity and remaining in the vicinity of the least
squares approximation, we anticipate
\[ | \tau - 1 | \ll 1 \spc \Longrightarrow \spc \mbox {a unique
solution} . \]

\item However, we have no clear indication yet on what occurs
further away
\[ | \tau - 1 | \gg \epsilon \spc \Longrightarrow \spc \mbox {any
number of solutions} . \]

\end{itemize}

\itt{Some numerical observations.} A numerical example let see
how the approximations $ \apx$ vary as a function of $ \tau$.

We search for the (unique or multiple) rank-1 approximations to
matrix $X$ under weights $ \left(w_{i, j} \right) $ in minimization
\rf{eq1},
\beq {eq8} X = \left(
\begin {array} {cc} 6 & 0 \\
1 & 2 \\
\end {array} \right) \et {\cal Z}_0 = \left(w_{i, j}^2 \right)
= \left(
\begin {array} {cc} 0.04 & 0.68 \\
0.84 & 0.40 \\
\end {array} \right) . \eeq
and it turns out that the data set \rf{eq8} has 2 rank-1 approximations
under weighted Frobenius norm \rf{eq1},
\skp{TREAT_M_N_P_SD (2,2,1, 950, 1=1);}
\[ \apxx {b,} 0 = \left(
\begin {array} {cc} 5.871 & 0.202 \\
1.032 & 0.036 \\
\end {array} \right), \]
and
\[ \apxx {c,} 0 = \left(
\begin {array} {cc} 0.101 & 0.194 \\
1.030 & 1.968 \\
\end {array} \right) . \]
-- Further on, the subscripts $b$ and $c$ will appear at Table
1 -- These two approximations clearly are very different and they
also are of different qualities. Defining a weighted root mean
square error, $\rms$, by
\[ \rms^2 = \frac {\sum_{i, j} w_{i, j}^2 \left(\apx_{i, j}
- X_{i, j} \right)^2} {\sum_{i, j} w_{i, j}^2}, \]
they respectively yield
\[ \rms_b = 0.8958 \et \rms_c = 0.8507. \]

The set of weights \rf{eq8} has the average
\[ \bar z = \frac {\sum_{i, j} w_{i, j}^4} {\sum_{i, j} w_
{i, j}^2} = 0.67837 \]
and the path $ {\cal Z}_ {\tau} $ is defined by Rule \rf{eq7},
the set of pseudo-weights linearly passing through the two sets
$ {\cal Z}_0 $ and $ {\cal Z}_1 = $ {\footnotesize $\left(
\begin {array} {rr} 0.67837 & 0.67837 \\
0.67837 &0.67837 \\
\end {array} \right) $}.

The solutions to saddle point problem \rf{eq6} vary along this
path. Each approximation can be seen as a point of a 1-dimension
``Curve" parametrized by $ \tau$ in the $ (m \times n) $-dimension
space of rank-$p$ approximations. For instance, the curve corresponding
to the \svd-approximation runs through the points
\[ \apxx {} {0.9} = \left(
\begin {array} {cc} 5.978 & 0.335 \\
1.099 & 0.062 \\
\end {array} \right), \spc \mbox {with} \; \rms = 0.9000, \]
\[ \apxx {} 1 = \left(
\begin {array} {cc} 5.978 & 0.351 \\
1.110 & 0.065 \\
\end {array} \right) = \svd, \spc \mbox {with} \; \rms = 0.9011,
\]
and
\[ \apxx {} {1.1} = \left(
\begin {array} {cc} 5.978 & 0.373 \\
1.125 & 0.070 \\
\end {array} \right) \spc \mbox {with} \; \rms = 0.9028. \]

This curve starts at $ \tau = - 0.05227 $ and terminates at $
\tau = 5.19696 $. The right end point of this curve will be referred
to as being a ``Cut"; there is a $w_{i, j}^2 (\tau) $ that approximately
cancels at this curve end, $ \left({\cal Z}_ {5.19697} \right)
_{2, 1} = 0 $. This curve also passes through the worst of the
two solutions, $ \apxx {b,} 0 $.

A thorough space search led to the finding of four different curves.
They are briefly described at Table 1.
\skp{

Curve_1. 1: -0.05227 -1.9406 0.2554 0.0498 \\
Nearest Cut (1,1) = -0.06266 \\
Curve_1. 51: 5.19696 -0.0037 0.1487 79.545 \\
Nearest Cut (2,1) = 5.19697 \\

Curve_2. 1: -0.18799 -6.0495 -0.5564 0.2014 \\
Curve_2. 51: 0.09357 -4.8576 0.6682 0.2862 \\

Curve_3. 1: -8.27280 -2.6435 -2.2306 0.8546 \\
Curve_3. 51: -0.06485 -1.9794 0.1512 0.0180 \\
Nearest Cut (1,1) = -0.06266 \\

Curve_4. 1: 4.07575 -1.7798 -0.9407 -2.3810 \\
Curve_4. 51: 5.19696 -0.0037 -0.1496 -81.044 \\
Nearest Cut (2,1) = 5.19697}

\btb {Four curves and the cuts of example \rf{eq8}.} {t1} {| c
| rr | c |}
\hline
Subscript & $ \tau (\mbox {\footnotesize Left end}) $ & $ \tau
(\mbox {\footnotesize Right end}) $ & Special $ \tau$-values \\
\hline
& & & -1.43695 $ = \cut 2 2 $ \\
a & -8.27280 & -0.06485 & \\
& & & -0.06266 $ = \cut 1 1 $ \\
b & -0.18799 & 0.09357 & 0 \\
c & -0.05227 & 5.19696 & 0 and 1 \\
d & 4.07575 & 5.19696 & \\
& & & 5.19697 $ = \cut 2 1 $ \\
& & & 416.500 $ = \cut 1 2 $ \\
\hline
\etb

This numerical example is small-size and bigger matrices are worthy
of attention. We now search for the (unique or multiple) rank-2
approximations to the next matrix $X$ under weights $ \left(w_
{i, j} \right) $ in minimization \rf{eq1},
\skp{TREAT_M_N_P_SD (4, 3, 2, 1817, 1 = 1);}
\beq {eq9} X = \left(
\begin {array} {ccc} 6 & 4 & 6 \\
2 & 2 & 9 \\
9 & 0 & 7 \\
1 & 3 & 1 \\
\end {array} \right) \et {\cal Z}_0 = \left(w_{i, j}^2 \right)
= \left(
\begin {array} {ccc} 0.04 & 0.84 & 0.72 \\
0.56 & 1 & 0.68 \\
0.12 & 0.40 & 0.52 \\
0.60 & 0.48 & 0.32 \\
\end {array} \right) . \eeq
Example \rf{eq9} has 3 solutions, namely the approximations
\[ \left(
\begin {array} {ccc} 9.372 & 3.431 & 6.079 \\
2.152 & 1.704 & 9.052 \\
6.448 & 2.668 & 6.754 \\
1.550 & 0.618 & 1.427 \\
\end {array} \right), \spc \left(
\begin {array} {ccc} 2.114 & 3.974 & 6.095 \\
3.424 & 2.112 & 8.486 \\
3.355 & - 0.239 & 7.572 \\
0.290 & 2.875 & 1.584 \\
\end {array} \right) \]
and
\[ \left(
\begin {array} {ccc} - 0.065 & 3.563 & 6.285 \\
2.908 & 2.774 & 8.363 \\
7.371 & - 0.743 & 7.320 \\
0.104 & 1.295 & 2.433 \\
\end {array} \right) . \]

As for example \rf{eq8}, we report the main observations at Table
2.

\btb {Curves and cuts of example \rf{eq9}.} {t2} {|c | rr | c
|}
\hline
Subscript & $ \tau (\mbox {\footnotesize Left end}) $ & $ \tau
(\mbox {\footnotesize Right end}) $ & Special $ \tau$-values \\
\hline
& & & -10.1509 $ = \cut 4 1$ \\
& & & -5.65039 $ = \cut 2 1$ \\
& & & -3.73810 $ = \cut 3 3$ \\
e & -3.36121 & -3.08193 & \\
& & & -2.67994 $ = \cut 4 2$ \\
f & -2.15933 & -2.10259 & \\
g & -1.82206 & -1.01699 & \\
& & & -1.54376 $ = \cut 3 2$ \\
& & & -0.94365 $ = \cut 4 3$ \\
h & -0.94702 & -0.32216 & \\
i & -0.24797 & 0.01177 & 0 \\
& & & -0.22259 $ = \cut 3 1$ \\
j & -0.06533 & 0.27532 &0 \\
& & & -0.06461 $ = \cut 1 1$ \\
k & -0.06461 & 2.93359 & 0 and 1 \\
l & 2.55480 & 2.93348 & \\
& & & 2.93348 $ = \cut 2 2$ \\
m & 3.75504 & 8.40023 & \\
& & & 4.64366 $ = \cut 1 2$ \\
n & 5.07162 & 6.66613 & \\
o & 8.51944 & $ > $ 20.00 & \\
& & & 11.8243 $ = \cut 1 3$ \\
p & 16.0803 & $ > $ 20.00 & \\
& & & 32.5488 $ = \cut 2 3$ \\
\hline
\etb

\itt{Discussion of numerical findings} The tables must be read
keeping in mind that the reported values of curve end-points have
a limited precision. They have been identified by following the
curves until they terminate and this procedure is somewhat coarse.

The two cases reported for illustration are somewhat different.
Example \rf{eq8} is so small that that the sheer appearance of
a zero weight yields to a rank reduction such that the `approximations'
realise exact fits (at $p = 1$). This is not the case for $\min
{m, n} > 1$ as in Example \rf{eq9}.

Let us now list our main observations. When referring to given
curves of Tables 1 and 2, we use the subscripts specified in those
two Tables.

\begin{enumerate}

\item Cuts play a crucial role as end-points of curves (see $a,
c, d, k, l$).

\item Most end-points are not in the vicinity of cuts.

\item Curves corresponding to paths \rf{eq7} may pass through
cuts (see $g, i, j, m, o$).

\item For $p > 1$, we observe that conditions \rf{eq5} are usually
satisfied at curve ends.

\item There are vicinities in $\tau$-values where no curve seem
to exist. They correspond to combinations of positive and negative
pseudo-weights.

\item The best solution at $\tau = 0$ cannot always be attained
from the SVD-solution at $\tau = 1$ (no curve may join these two
solutions, see $b$ and $c$ of Table 1).

\item The number of solutions at $\tau = 0$ is at least 1 and
at most the minimum dimension of the approximated matrix. \\
Table 3 reports the findings based on a large population of $
(X, {\cal Z}_0) $-pairs.

\btb {Maximum number of solutions.} {t3} {| c c | c |c|}
\hline
\multicolumn{2}{|c|}{Dimensions} & Approximation & Maximum \\
$m$ & $n$ & rank & number of solutions \\
\hline
2 & 2 & 1 & 2 \\
2 & 3 & 1 & 2 \\
2 & 4 & 1 & 2 \\
2 & 5 & 1 & 2 \\
2 & 6 & 1 & 2 \\
3 & 2 & 1 & 2 \\
4 & 2 & 1 & 2 \\
5 & 2 & 1 & 2 \\
6 & 2 & 1 & 2 \\
3 & 3 & 1 & 3 \\
3 & 4 & 1 & 3 \\
3 & 4 & 2 & 3 \\
3 & 5 & 1 & 3 \\
3 & 5 & 2 & 3 \\
3 & 6 & 1 & 3 \\
4 & 3 & 1 & 3 \\
4 & 3 & 2 & 3 \\
5 & 3 & 1 & 3 \\
5 & 3 & 2 & 3 \\
6 & 3 & 1 & 3 \\
4 & 4 & 2 & 4 \\
4 & 5 & 2 & 4 \\
4 & 5 & 3 & 4 \\
4 & 6 & 2 & 4 \\
4 & 6 & 3 & 4 \\
5 & 4 & 2 & 4 \\
5 & 4 & 3 & 4 \\
6 & 4 & 2 & 4 \\
6 & 4 & 3 & 4 \\
5 & 6 & 2 & 5 \\
5 & 6 & 3 & 5 \\
5 & 6 & 4 & 5 \\
6 & 5 & 2 & 5 \\
6 & 5 & 3 & 5 \\
6 & 5 & 4 & 5 \\
\hline
\etb

\end{enumerate}

\itt{A conjecture} {\em Given a $m \times n$-matrix $X$, it has
up to $\miin {} (m, n) $ weighted low-rank approximations.} This
is observed in the context of weighted Frobenius norm \rf{eq1}.

\itt{Algorithmic notes} Two main numerical difficulties are encountered
and both are approached by the use of ``closest bases". We first
introduce this concept and then describe the two difficulties,
namely the path following procedure and the thorough space search.

\begin{itemize}

\item Closest basis.

Given a set of vectors $a_1, ..., a_p$, the Gram-Schmidt process
is standard to build up a basis that is normed and orthogonal
and that permits a perfect decomposition of these original vectors.
This Gram-Schmidt basis is clearly not unique, seeing that different
sets of such bases can be constructed by simply permuting the
original vectors; unfortunately, this lack of uniqueness seriously
complicate the numerical steps.

The classical Gram-Schmidt process is now reminded by \rf{eq10},
before being slightly modified into \rf{eq11}.
\beq {eq10}
\begin {array} l \mbox {\small for} \; i : = 1 \; \mbox {\small
to} \; p \; \mbox {\small do} \\
\spc \left|
\begin {array} {ll} e_i : = a_i & \; \mbox {\small Base vector
initialisation.} \\
\delta_i : = \sum_{j < i} \; (e_j'\; e_i) \; e_j & \; \mbox {\small
Projection on previous $e_j$.} \\
e_i : = e_i - \delta_i & \; \mbox {\small Deflation.} \\
e_i : = e_i / (e_i'\; e_i)^ {1 / 2} & \; \mbox {\small Normalisation.}
\\
a_i : = e_i & \; \mbox {\small Substitution.}
\end {array} \right.
\skp {. | close}
\end {array} \eeq
The stability can be greatly improved by constructing a basis
that is ``close" in directions to the original set $a_1, ...,
a_p$.

The ``closest basis" is derived according to the next algorithm,
an original method as far as we know.
\beq {eq11}
\begin {array} l \mbox {\small Repeat} \\
\spc \left|
\begin {array} {ll} \mbox {\small for} \; i : = 1 \; \mbox {\small
to} \; p \; \mbox {\small do} & \; \mbox {\small Internal loop
1.} \\
\spc e_i : = a_i / (a_i'\; a_i)^ {1 / 2} & \; \mbox {\small Initialisation
and normalisation.} \\
\mbox {\small for} \; i : = 1 \; \mbox {\small to} \; p \; \mbox
{\small do} & \; \mbox {\small Internal loop 2.} \\
\spc \delta_i : = \frac 1 2 \; \sum_{j \neq i} \; (e_j'\; e_i)
\; e_j & \; \mbox {\small Half projection on all others.} \\
\mbox {\small for} \; i : = 1 \; \mbox {\small to} \; p \; \mbox
{\small do} & \; \mbox {\small Internal loop 3.} \\
\spc e_i : = e_i - \delta_i & \; \mbox {\small Deflation.} \\
\spc a_i : = e_i & \; \mbox {\small Substitution.} \\
\end {array} \right.
\skp {. | close} \\
\mbox {\small until sufficient convergence.}
\end {array} \eeq
Algorithm \rf{eq11} is iterative and therefore is slower than
the Gram-Schmidt standard by \rf{eq10}. Its convergence is quadratic
and the global loop is little run. The extreme stability of this
``closest basis" justifies the expense.

\item Path following.

The path $ {\cal Z}_ {\tau} $ by \rf{eq7} is with respect to
the pseudo-weights and is associated to varying approximations
\[ {\cal Z}_ {\tau} = {\cal Z}_0 + \tau \; ({\cal Z}_1 - {\cal
Z}_0) \spc \rightarrow \spc \apx_{\tau} = (A \; B')_ {\tau}
= A_{\tau} \; B'_{\tau} . \]
We remark that, even if $ (A \; B')_ {\tau} $ is defined, $A_
{\tau} $ remains indeterminate seeing \rf{eq4}. However, if we
knew $A_{\tau} $, $B_{\tau} $ would immediately result from
\rf{eq2},
\[ B_{\tau} : \; \mnz {X - A_{\tau} \; B'}_ {w^2_{\tau}}^2
= \miin B. \]

In order to follow the path \rf{eq7}, we impose a smooth variation
on $A_{\tau} $. On the one hand, we restrict $A_{\tau} $ to
be orthonormed and, on the other hand, to be slowly varying.

Given a $A_{\tau} $ solution of \rf{eq6} and corresponding to
a given $\tau$-value, its $p$ column vectors $a_1, ..., a_p$ have
size $m$, $m > p$, and we orthonormalise by algorithm \rf{eq11}.
This is our first solution. \\
Slightly modifying the first $\tau$-value and with the help of
the first solution as initialisation, we derive a new solution
$A_{\tau} $. This is our second solution and it lies in a tight
vicinity of the first (due to the fact that we used a closest
basis rather than the classical Gram-Schmidt process). We have
obtained two points of a curve. \\
Further on, we apply a predictor-corrector algorithm with step
length adaptation. Several strategies of Bates et al. (2008) are
relevant.

\item Thorough space search.

Counting the number of solutions existing for a given $\tau$-value
is quite a problem. We resorted to applying a space search strategy.
\\
We start with an arbitrary initialisation and solve minimization
\rf{eq1}. This yields a given solution $ \apx_{\tau} $ (at $\tau
= 0$). Repeating with other initialisations, we obtain either
new $ \apx_{\tau} $ or repeats of the already known solutions.

The main trouble with the above strategy is due to possibly small
radii of convergence. Some solutions $ \apx_{\tau} $ can be discovered
only when an initialisation is performed in their immediate vicinity;
starting too far away, the minimizations converge toward some
``dominant" solutions.

Hence, it is peremptory to apply initialisations which evenly
span the search space.

Initialising $N$ times, we need $N$ good starting $A_{\tau} $,
each with $m\times p$ entries. These entries will be the coordinates
of $N$ points on a ball in $ {\mathbb R}^ {m \times p} $; eventually,
each of the $A_{\tau} $-points is projected onto a subspace by
orthonormalisation into a closest basis. Gross (2011) and Recht
(2011) discuss what the sample size $N$ must be.

First, the $N$ points on the ball are assigned as being the summits
of a nearly regular polyhedron; then, they are slightly shifted
as if they were exerting a repulsive force on the other points.
We construct a dispersed phase of points on the ball surface.

\end{itemize}

\end{enumerate}
\[ \]

\mbox{} \\
{\bf \sf References} \\
{\small \mbox{} \\
Bates D.J., Hauenstein J.D., Sommese A.J. and Wampler C.W (2008).
Adaptive Multiprecision Path Tracking, {\em SIAM J. Numer. Anal.},
46, 722-746. \\
\mbox{} \\
Benzi M., Golub G.H. and Liesen J. (2005). Numerical solution
of saddle point problems, {\em Acta Numerica}, 14, 1 - 137 \\
\mbox{} \\
Bojanczyk A., Higham N.J. and Patel H. (2003). Solving the indefinite
least squares problem by hyperbolic QR factorization, {\em SIAM
J. Matrix Anal. Appl.}, 24, 914 - 931. \\
\mbox{} \\
Bradu, D. and Gabriel, K.R. (1978). The biplot as a diagnostic
tool for model of twoway tables, {\em Technometrics}, 20, 47-68.
\\
\mbox{} \\
Gabriel K.R. and Zamir S. (1979). Lower Rank Approximation of
Matrices by Least Squares with Any Choice of Weights, {\em Technometrics},
21, 489-498. \\
\mbox{} \\
Gillis N. and Glineur F. (2010). Low-rank matrix approximation
with weights or missing data is NP-hard, {\em http://arxiv.org/abs/1012.0197}.
\\
\mbox{} \\
Greenacre M.J. (2012) Biplots: the joy of singular value decomposition,
{\em Wiley Interdisciplinary Reviews: Computational Statistics},
4, 399-406. \\
\mbox{} \\
Gross D. (2011) Recovering low-rank matrices from few Coefficients
in any basis {\em IEEE Transactions on Information Theory}, 57,
1548 - 1566. \\
\mbox{} \\
Higham, N.J. (2002) Computing the nearest correlation matrix,
a problem from finance, {\em IMA Journal of Numerical Analysis},
22, 329-343. \\
\mbox{} \\
Manton J.H., Mahony, R. and Hua, Y. (2003). The geometry of weighted
low-rank approximations. {\em IEEE Transactions on Signal Processing},
51, 500-514. \\
\mbox{} \\
Markovsky I. (2010). Algorithms and literate programs for weighted
low-rank approximation with missing data, Preprint. \\
\mbox{} \\
Markovsky I. and Van Huffel S. (2007). Left vs right representations
for solving weighted low-rank approximation problems. {\em Linear
Algebra and its Applications}, 422, 540-552. \\
\mbox{} \\
Maronna R.A. and Yohai V.J. (2008). Robust lower-rank approximation
of data matrices with element-wise contamination. {\em Technometrics},
50, 295-304. \\
\mbox{} \\
Okatani T., Yoshida T. and Deguchi, K. (2011). Efficient algorithm
for low-rank matrix factorization with missing components and
performance comparison of latest algorithms. {\em 2011 IEEE International
Conference on Computer Vision (ICCV)}, 6-13 Nov. 2011, 842-849.
\\
\mbox{} \\
Recht B. (2011) A simpler approach to matrix completion, {\em
Journal of Machine Learning Research}, 12, 3413-3430 \\
\mbox{} \\
Simonsson L. and Eld\'en L. (2010). Grassmann algorithms for low-rank
approximation of matrices with missing values, {\em BIT Numerical
Mathematics}, 50, 173-191. \\
\mbox{} \\
Srebro N. and Jaakkola T. (2003). Weighted low-rank approximations.
In {\em ICML, 20th International Conference on Machine Learning},
720-727. \\
\mbox{} \\
Usevich K. and Markovsky I. (2012) Variable projection for affinely
structured low-rank approximation in weighted 2-norm. {\em http://arxiv.org/pdf/1211.3938.}
\\
\mbox{} \\
Yan G. (2010). Structured low-rank Matrix Optimization Problems:
A Penalty Approach, {\em Thesis}.}

\end{document}